\documentclass[prd,aps,twocolumn,amsmath,amssymb,nofootinbib,preprintnumbers]
{revtex4}

\pdfoutput=1
\usepackage{graphicx}
\usepackage{dcolumn}
\usepackage{bm}
\usepackage{amsmath}
\usepackage{amsfonts}
\usepackage{euscript,bbm}
\usepackage{ifthen}
\usepackage{psfrag}

\def\ls{\mathrel{\lower4pt\vbox{\lineskip=0pt\baselineskip=0pt
           \hbox{$<$}\hbox{$\sim$}}}}
\def\gs{\mathrel{\lower4pt\vbox{\lineskip=0pt\baselineskip=0pt
           \hbox{$>$}\hbox{$\sim$}}}}
\def\drawbox#1#2{\hrule height#2pt

\hbox{\vrule width#2pt height#1pt \kern#1pt
              \vrule width#2pt}
              \hrule height#2pt}

\def\Asym#1#2{\vcenter{\vbox{\drawbox{#1}{#2}
              \kern-#2pt       
              \drawbox{#1}{#2}}}}

\newcommand{\PRE}[1]{}
\newcommand{\Expect}[1]{\left\langle #1 \right\rangle}
\newcommand{\beq}{\begin{equation}}
\newcommand{\eeq}{\end{equation}}

\begin{document}

\preprint{MIFP-09-19, NSF-KITP-09-52, arXiv: 0904.3773}
%
\title{Mirage in the Sky: Non-thermal Dark Matter, Gravitino Problem, and Cosmic Ray Anomalies}

\author{Bhaskar Dutta$^1$}
\email{dutta@physics.tamu.edu}
\author{Louis Leblond$^{1,2}$}
\email{lleblond@physics.tamu.edu}
\author{Kuver Sinha$^1$}
\email{kuver@physics.tamu.edu}
\affiliation{ $^1$ George P. \& Cynthia W. Mitchell Institute for Fundamental Physics,
  Texas A\&M University, College Station, TX 77843-4242\\
$^2$ Kavli Institute for Theoretical Physics, University of California, Santa Barbara, CA,
93106, USA
}




\begin{abstract}
Recent anomalies in cosmic rays could be due to dark matter annihilation in our galaxy. In order to get the required large cross-section to explain the data while still obtaining the right relic density, we rely on a non standard thermal history between dark matter freeze-out and Big-Bang Nucleosynthesis (BBN). We show that through a reheating phase from the decay of a heavy moduli or even the gravitino, we can produce the right relic density of dark matter if its self-annihilation cross-section is large enough. In addition to fitting the recent data, this scenario solves the cosmological moduli and gravitino problems. We illustrate this mechanism with a specific example in the context of $U(1)_{B-L}$ extended MSSM where supersymmetry is broken via mirage mediation. These string motivated models naturally contain heavy moduli decaying to the gravitino, whose subsequent decay to the LSP can reheat the universe at a low temperature. The right-handed sneutrino and the $B-L$ gaugino can both be viable dark matter candidates with large cross-section. They are leptophilic because of $B-L$ charges. We also show that it is possible to distinguish the non-thermal from the thermal scenario (using Sommerfeld enhancement) in direct detection experiments for certain regions of parameter space.
\end{abstract}

\maketitle


\section{Introduction}

Recent data from PAMELA shows an excess of positrons at energies in the 10-100~GeV range~\cite{PAMELA}, with data expected up to $\sim 270$~ GeV. No excess of anti-proton flux is observed~\cite{PAMELA-antiproton}. There is also new data from ATIC where one observes excess in $e^{+} + e^{-}$ spectrum with a peak around 600~GeV~\cite{ATIC}. PPB-BETS~\cite{pep}, also reports excess in the $e^+ + e^-$ energy spectrum between 500-800 GeV. While there could be astrophysical explanations for these anomalies~\cite{pulsars}, it is also possible that these are among the first signals of dark matter annihilation.

If PAMELA is explained by a WIMP dark matter, the data leads us to the following three broad characteristics for this particle:  It must be heavier than $\sim 100$  GeV, it must be leptophilic and it must have a large cross-section today~\cite{Vernon}. The first property is needed to explain the high energy positron detected while we need to have final states of dark matter annihilation  predominantly to be leptons in order to not overproduce anti-protons \cite{Strumia,Salati}.  Both of these properties can be easily arranged in a model dependent way. The necessarily large cross-section on the other hand is harder to fiddle with since it is directly constrained by the relic density. The thermally produced relic density is given by
\beq\label{thermal}
\Omega_{CDM} = 0.23 \left({3\times 10^{-26} cm^3 s^{-1} \over\Expect{\sigma v} }\right)\; .
\eeq
An interesting proposal is to enhance the cross-section at low-velocity through the Sommerfeld effect~\cite{Sommerfeld} whereas a light boson provides an attractive potential that enhances the cross-section when the dark matter is non-relativistic~\cite{Strumia,Nima} (also see~\cite{Allahverdi:2008jm,Other} for explanations using Sommerfield effect). In order to generate the right enhancement factor one needs to fix the ratio of the dark matter mass and the new light boson mass (or different parameters of the model) to a high degree of accuracy.

In this paper, we will be interested in a second alternative where we have a non-standard thermal history and Eq.~(\ref{thermal}) is modified. This possibility has already been explored as a possible explanation for PAMELA in \cite{Fairbairn:2008fb, Grajek} (see also \cite{Kane:2008gb}).  Such a non standard thermal history is very well-motivated from physics beyond the standard model and, of course, we have no direct evidence that the universe is radiation dominated at temperature above the Big-Bang Nucleosynthesis (BBN) temperature. If there is a phase of matter or dark energy domination prior to BBN, the only way to connect to radiation (whose energy density decays faster than either) is through a reheating process~\cite{Giudice}. If the reheating temperature ($T_r$) is in between the freeze-out temperature ($T_f$) and the BBN temperature, we will respect all current astrophysical constraints while the entropy produced at reheating will naturally dilute the relic density produced at freeze-out. If dark matter is produced non-thermally at the time of reheating, larger annihilation cross-sections are needed to obtain the right relic density. Similarly, we could have a phase dominated by a fluid whose energy density decays faster than radiation (e.g. kination~\cite{Salati:2002md}). In this case, there is no reheating phase but the Hubble expansion is faster than usual and the dark matter candidate must have a stronger coupling in order to get the right relic density. We therefore see that \emph{almost any} thermal history other than radiation domination would require a dark matter candidate with larger annihilation cross-section to obtain the correct relic density.

To illustrate this non-standard thermal history, we work with the specific example of a phase of matter domination before BBN. We consider the case where the matter component is a scalar field coherently oscillating (a cosmological modulus) and also the case where the matter component is the gravitino. The former case is well motivated from string theory where there are many flat directions (moduli) that acquire masses from supersymmetry breaking.

A second purpose of this paper is to explore the implications of such cosmological enhancement of the dark matter annihilation cross section on the cosmological moduli and gravitino problems. Generic string moduli of masses around the electroweak scale decay (and reheat the universe) after  BBN, ruining its successful predictions. A standard solution to this problem is to take the moduli to be heavy (at least 20 TeV), thus enabling them to decay before BBN. The decay of such moduli primarily produces gauge bosons, gauginos and dark matter; however, there is also gravitini production which is generically unsuppressed, with a branching ratio of around $0.01$ \cite{Endo} (this can be avoided in special set-ups \cite{Dine}). The production of gravitinos is again problematic and they need to be heavy enough to decay before  BBN. Moreover, the gravitinos decay to the LSP, and there are strong bounds to avoid overproduction of dark matter. Avoiding overproduction can cause the gravitino mass to be high (around $1000$ TeV) if the annihilation cross section of LSP is at the canonical value. This has disastrous implications for low-energy supersymmetric model building. Superconformal anomaly mediated contributions to the soft masses push the low energy superparticle spectrum into the $100$ TeV region.

We will show that a large enhancement of the cross-section ($\sim 10^3$) and heavier dark matter (TeV scale) can naturally ease the bounds on gravitino mass coming from overproduction of LSP. The gravitino becomes a legitimate candidate (together with cosmological moduli) to act as the decaying particle that non-thermally produces the right relic density of LSP. We thus solve the gravitino problem by having it to decay prior to  BBN, while in order not to overproduce the LSP we need the large cross-section required by PAMELA!

Non-thermal dark matter production and the moduli and gravitino problems have usually been studied in the context of a Wino LSP, which arises in models of anomaly mediation \cite{Moroi}, simple realizations of split supersymmetry, and in the context of the $G_2$-MSSM \cite{Acharya}.

In view of recent data, the high energy positron excess reported by PAMELA is difficult to fit with a LSP in the Wino mass range, unless non-standard assumptions are made for the distribution of dark matter and the propagation of cosmic rays (this has been studied extensively in ~\cite{Grajek}). Moreover, while it is possible that the anti-proton data suffers from theoretical uncertainties in cosmic ray propagation, taken at face value such data appears not to prefer a Wino LSP.

The point of view we will take in this paper is that the above cosmological scenario can work in a $U(1)_{B-L}$ extension of the MSSM, in the setting of mirage mediation. From a model-building perspective, the fact that a non-minimal model eases bounds is perhaps not entirely surprising. However, as we will show, this particular extension (already well-motivated by non-zero neutrino mass) has a TeV-scale leptophilic LSP (the right-handed sneutrino or the $B-L$ gaugino) with large cross section.  The string inspired models of mirage mediation also solve the tachyonic slepton problem of anomaly mediation.

In comparison to the usual thermal production of dark matter, an enhancement factor given by the ratio of the freeze-out temperature to the reheat temperature gets generated in this scenario. After solving the moduli and the gravition problem, we will show that the enhancement factor is of order $\frac{T_f}{T_r} \sim 10^{3-4}$ which is in the right range to explain the cosmic ray puzzle.

We note that for mirage mediation in the MSSM, the LSP is primarily the Bino, which is unacceptable in light of the PAMELA data while our dark matter candidates in the $U(1)_{B-L}$ extension can fare better. Thus, apart from solving the moduli/gravitino problem, our model connects (possible) indirect observation of dark matter with string inspired phenomenology.  We also show that in the case of a right-handed sneutrino, it is possible to distinguish the non-thermal from the thermal scenarios in direct detection experiments.

In section \S 2 we give details of the cosmological enhancement, including the gravitino problem. In section \S 3, we work out the example of the mirage mediated $B-L$ extension, which provides a concrete model where the above cosmological history can occur. We conclude in section \S 4.

\section{Cosmological Enhancement}

Here we will work out the cosmological enhancement of the cross section caused by the reheating of a heavy modulus before BBN (the story is similar for the gravitino decay). The key phenomenon here is the low temperature of reheat (LTR) \cite{Giudice} which could also come from a phase of dark energy dominance such as low scale inflation or thermal inflation instead of a phase dominated by matter. Cosmological moduli can come to dominate the energy density of the universe if they are displaced from the minimum of their potential.
The equation of motion of a scalar field with gravitational strength decay rate in a FRW background is
\beq
\ddot\phi + (3H + \Gamma_\phi) \dot\phi + V' = 0\; .
\eeq
After inflation, the initial vev of the field ($\phi_{in}$) is displaced from its zero temperature minimum by some amount $M$ (say inflationary scale). At early time, $H > m_\phi$, the friction term dominates over the potential and the field is frozen at its initial value $\phi =\phi_{in}$. The universe is then radiation dominated until $t \sim m_\phi^{-1}$ (at a temperature $T_{in} \sim \sqrt{m_\phi M_p}$) at which point the field will start oscillating around its minimum. These coherent oscillations of a scalar field lead to large occupation numbers and the end result is a Bose-Einstein condensate which behaves like matter. The initial energy density ($m_\phi^2\phi_{in}^2$) will increase compared to radiation and it will eventually dominate until the modulus decays and reheats. If the modulus decays after BBN, the energy released will photo-dissociate the newly built nucleus \cite{Kawasaki} which is the crux of the cosmological moduli problem. In the following, we will take
  the cross-section of the modulus to ordinary matter and to dark matter to be Planck suppressed
\beq
\Gamma_{\phi} = \frac{c}{2\pi} \frac{m_\phi^3}{M_p}\; ,
\eeq
where we take $c\sim 1$ and $M_p = 2.4 \times 10^{18}$GeV is the reduced Planck mass. In the approximation of sudden decay the reheating temperature can be defined by taking the lifetime of the modulus ($\Gamma_\phi^{-1}$)  to be equal to the expansion rate at the time of reheating $t = \frac{2}{3 H}$. Right after reheating (at $T = T_r$) the universe is radiation dominated with $H  = \sqrt{\frac{\pi^2g_*}{90}}{T_r^2 \over M_p}$  where $g_*$ counts the number of degrees of freedom\footnote{This relation is modified by $\mathcal{O}(1)$ factor because of the non-standard cosmology \cite{Giudice} with the massive modulus.}. Since our temperature of reheat will be always be around $10 - 100$ MeV, $g_*$ has the usual value of $g_* = 10.75$. The temperature of reheat is then
\begin{eqnarray}\label{Tr}
T_r &\approx &\sqrt{\Gamma M_p} \sim \frac{c^{1/2} m_\phi^{3/2}}{M_p}\; ,\nonumber \\
& = &c^{1/2} \left( \frac{10.75 }{ g_*}\right)^{1/4}
\left( \frac{m_\phi}{100\, {\rm TeV}}\right)^{3/2}\, 6.37\, {\rm MeV}.
\end{eqnarray}
We can compute the relic density of dark matter using the Boltzmann equations for the modulus $\phi$, the dark matter candidate $X$ and radiation $R$ \cite{Moroi}
\begin{eqnarray}
\frac{d \rho_\phi}{dt} &=& -3H \rho_\phi-\Gamma_\phi \rho_\phi
\label{binp} \;,  \\
\frac{d \rho_R}{dt} &=& -4H \rho_R + (m_\phi - N_{LSP}m_X) \Gamma_\phi n_\phi \nonumber\\
&& + \langle \sigma v \rangle 2 m_{X} \left[ n_{X}^2 - \left({n_{X}^{eq}}\right)^2\right]
\label{binr} \;, \\
\frac{d n_{X}}{dt} &=& -3H n_{X} + N_{LSP} \Gamma_\phi n_X \nonumber\\
&&-\langle \sigma v \rangle  \left[ n_{X}^2 -
\left({n_{X}^{eq}}\right)^2\right]
\end{eqnarray}
where $N_{LSP}$ is the average number of LSP particles produced by the decay of one modulus and we have taken the energy of both $\phi$ and $X$ to be simply given by their masses, i.e. they are non-relativistic.
Different initial conditions at freeze-out are possible and we will look at the case where the universe is dominated by the moduli prior to freeze-out\footnote{Demanding that the universe is matter dominated prior to freeze-out imposes the following condition on the initial value of $\phi$, $\phi_{in}> T_f^{1/2}\frac{M_p^{3/4}}{m_\phi^{1/4}}$ which is around $10^{(13-14)}$ GeV for the numbers considered in this paper. This is well below the expected value of $\phi_{in} \sim M_p$.}. For high masses and strong enough interactions, the dark matter candidate will be non-relativistic at the time of freeze-out (with $n_{X}^{eq}= g_* \left(\frac{M_X T}{2\pi}\right)^{3/2} e^{-M_x/T}$) and it will have time to reach equilibrium before reheating occurs. Dark matter freeze-out occur when the annihilation rate is equal to the rate of expansion
\beq\label{freezeout}
\Gamma_X = n_X^{eq}(T_f) \Expect{\sigma v} = H(T_f)
\eeq
up to the fact that the thermal history is changed due to the presence of the decaying moduli. This change is relatively minor and the new freeze-out temperature $T_f^{new}$ is very close to the usual one at $T_f^{old}  \sim m_{X}/20$ which is what we will be using.
After freeze-out, reheating occurs and the entropy production will dilute the initial density of dark matter by a total factor of $T_{r}^3/T_f^3$ which can  be as much as $10^{-12}$.  To first order, we can therefore neglect the initial density of dark matter coming from freeze-out and instead just focus on the component produced non-thermally from the decay of the heavy moduli.

To compute the non-thermally produced dark matter, we can use the fact that there is an attractor solution to the Boltzmann equations. The idea is that if dark matter is overproduced by the moduli (compared to what one has for a usual freeze-out) they will quickly annihilate back into radiation. Therefore the maximal density of dark matter is given by the same condition we used before (Eq.~\ref{freezeout}) but now at a lower temperature
\beq
n_X^{eq}(T_r) \Expect{\sigma v} = H(T_r)\; .
\eeq
The non-thermal density of dark matter scales like
\beq
\Omega_X^{NT}(T_r) \sim \frac{n_X (T_r)}{s(T_r)} \sim \frac{H(T_r)}{T_r^3} \sim \frac{1}{T_r}\; ,
\eeq
where $s = \frac{2\pi^2}{45} g_* T^3$ is the entropy density. This should be compared to the usual thermal freeze-out density
\beq
\Omega_X^{T}(T_r) \sim \frac{n_X (T_f)}{s(T_f)} \sim \frac{1}{T_f}\; ,
\eeq
Hence the non-thermal production is enhanced compared to the usual thermal one by a factor
\beq
\Omega_X^{NT} = \Omega_X^T \frac{T_f}{T_r} = 0.23 \left({3\times 10^{-26} cm^3 s^{-1} \over\Expect{\sigma v} }\right)\frac{T_f}{T_r}
\eeq
and we must increase the cross-section accordingly to explain the data (by the factor $T_f/T_r$).

In the case, where the modulus has a small branching ratio to the dark matter particle (small $N_{LSP}$) or in the case where the modulus does not dominate the energy density before it decays, one may not reach the attractor solution. In this case, one can show that the dark matter abundance (or yield) $Y_X(T) \equiv \frac{n_X}{s(T)}$ is just given by the abundance of the modulus at reheating times the branching ratio $B_{\phi \rightarrow X} Y_{\phi}$. If one work with number density instead of abundance, we should use the average number of particle produced $N_{LSP}$ instead of the branching ratio as in \cite{Moroi}.

Therefore the abundance of LSP is the minimum
\beq
Y_{X}(T_r) = \rm{min} \left( B_{\phi \rightarrow  X} Y_{\phi} (T_r)\; \; , \;\; \sqrt{\frac{45}{8\pi^2 g_*}}\frac{1}{M_p T_r\Expect{\sigma v}}\right)\; .
\eeq
The first possibility represents the case where not enough dark matter is produced for self annihilation to start while the second is the attractor solution described above. 
 At this point the mass of the modulus is a free parameter and we can tune it to get any temperature of reheat desired while tuning the cross-section accordingly to get the right relic density \footnote{Note that in addition to the BBN constraints, the moduli is also constrained by WMAP measurements of isocurvature perturbations \cite{Lemoine}.  We leave it to future work to check the implications of these constraints on non-thermal production of dark matter from moduli.  If dark matter is produced by the decay of a gravitino, instead of a modulus, no isocurvature perturbations will be produced.}. For a very heavy modulus, an important worry is that gravitino will be produced in the reheating phase creating a new problem (or rather reviving an old one) \cite{Endo}.

\subsection{Gravitino Decay}

Since the gravitino has Planck suppressed couplings, it is never in thermal equilibrium in the early universe and depending on how much of it is produced at  various reheating phases, it can come to dominate the energy density and ruin BBN just like the cosmological moduli can (in this paper we are assuming that the gravitino is not the LSP and that it  decays).

If this is the only reheating phase in the early universe, then the abundance is directly proportional to the reheating temperature and solving the Boltzmann equations for the inflaton/radiation/gravitino system gives
\beq
Y_{3/2} \sim 2\times 10^{-12} \left(\frac{T_r^{inf}}{10^{10} \rm{GeV}}\right)
\eeq
The BBN constraint on the temperature of reheat from inflation ($T_r^{inf}$) can be very stringent. For $m_{3/2}\sim 30$ TeV, $Y_{3/2}$ must be smaller than $2\times 10^{-12}$ at 95\% confidence level which implies $T_r^{inf} < 10^{10}$ GeV while for smaller values (say $m_{3/2} \sim$ TeV) the bounds are more severe and  $T_r^{inf}$ needs to be as low as $\sim 10^6$ Gev which is a very serious constraint on inflationary models \cite{Kohri}.

Assuming that the temperature of reheat from inflation satisfies the constraints, the gravitino problem is revived in the presence of decaying moduli and a LTR \cite{Endo}. Indeed assuming a branching fraction $B_{3/2}$ of the moduli $\phi$ to $\psi_{3/2}$, then from non-thermal production during the decay of $\phi$, we produce gravitini with
\begin{eqnarray}
Y_{3/2} & = & 2 B_{3/2} Y_\phi = \frac32 B_{3/2} \frac{T_r}{m_\phi}\; , \nonumber \\
& = & 8\times 10^{-8} c^{1/2} B_{3/2} \left(\frac{m_\phi}{10^2 \rm{TeV}}\right)^{\frac12}\; ,
\end{eqnarray}
where $Y_\phi = \frac{3T_r}{4m_\phi}$ can be obtained directly from the Boltzmann equation for $\phi$. Unless $B_{3/2}$ is tuned to be small, a low mass gravitino (again say less than 20 TeV) is ruled out in this scenario and the heavy moduli give rise to the gravitino problem.

One way out is to assume that the moduli has suppressed couplings to the gravitino and that $B_{3/2}$ is naturally small. On general grounds we expect this branching ratio to be of order $0.01 - 1$\cite{Endo} for $m_\phi \gg m_{3/2}$. In \cite{Dine}, it was argued that it could be smaller due to helicity suppression. In this case the decay width of the modulus to gravitino is suppressed from its total decay width $\Gamma_{total} \sim \frac{m_\phi^3}{m_p^2}$  to $\Gamma_{\phi\rightarrow\psi^{3/2}} \sim \frac{m_{3/2}^3}{M_p^2}$. Given a hierarchy between $m_\phi$ and $m_{3/2}$ of order of $4\pi^2$ (as an example), we can get a branching ratio of order $10^{(-4)} \sim 10^{(-5)}$ which can be enough to evade the BBN constraints. Alternatively, if $m_\phi \leq 2 m_{3/2}$ then the branching ratio is drastically reduced due to phase space consideration.

Given that the gravitino problem is so pervasive in many models of the early universe, it is tempting to assume that the abundance is not tuned (or diluted) to be small. The gravitino will dominate the energy density of the universe but if it decays and reheats prior to BBN, there will be no problem.
To get this one needs a fairly heavy gravitino and one is lead to SUSY breaking pattern of the type of anomaly mediation or mirage mediation. In this case there is hierarchy between the gravitino mass and the LSP which we will parametrize
\beq
m_{3/2} = \kappa m_X\; .
\eeq
In mirage mediation that we will discuss below, this hierarchy is of order $\kappa = 4\pi^2$.

Now the gravitino itself will produce dark matter with a yield of
\beq\label{Xfrom32}
Y_{X} = \rm{min} \left(Y_{3/2}\; \; , \;\; \sqrt{\frac{45}{8\pi^2 g_*}}\frac{1}{M_p T_{3/2}\Expect{\sigma v}}\right)
\eeq
where we have assumed a branching ratio of order 1 for gravitino decay into the LSP. The temperature of reheat for the gravitino $T_{3/2}$ is determined in the same way as for the cosmological moduli with a decay rate
\beq	
\Gamma_{3/2} = \frac{c_{3/2}}{2\pi} \frac{m_{3/2}^3}{M_p^2}\; .
\eeq
So $T_{3/2}$ is given by Eq.~(\ref{Tr}) replacing $c\rightarrow c_{3/2}$ and $m_\phi \rightarrow m_{3/2}$. In \cite{Endo}, it was shown that for MSSM dark matter candidates such as the Wino (with $\Expect{\sigma v} \sim 10^{-24}$), the gravitino will overproduce dark matter unless its mass is higher than around $\sim 10^3$ TeV. A similar problem was pointed out in Moroi and Randall \cite{Moroi} in the context of LSP production from modulus decay, where it was argued that a modulus mass of around $300$ TeV gave $\Omega_{LSP} \sim 1$, while a lower value of $\Omega \sim 0.1$ is obtained for even higher modulus mass.



Assuming that $Y_{3/2}$ is large, the abundance of dark matter is given by the second factor in Eq.~(\ref{Xfrom32}). The enhancement factor is given by
\beq \label{enhancement}
\frac{T_f}{T_{3/2}} = 6.14 \times 10^6 \left(\frac{\rm{TeV}}{m_X}\right)^{1/2} \frac{1}{\kappa^{3/2} c_{3/2}^{1/2}}\; .
\eeq
The branching ratio of the gravitino to LSP is essentially 1 and $c_{3/2}$ is constrained to be maximally around $1.5$ \cite{Moroi2} (essentially, supersymmetry fixes the coupling of the gravitino to the supercurrent).
For a hierarchy of $\kappa = 4\pi^2$ and a LSP mass around $m_X\sim 1.5$ \rm{TeV} an enhancement factor of $T_f/T_r \sim 10^4$ is obtained, which would require a cross-section of order $\Expect{\sigma v} \sim 10^{-22} cm^3 s^{-1}$. As we will discussed more in the subsequent sections,  it is possible to fit PAMELA with such a high cross-section although there is a definite tension with BBN constraints coming from dark matter annihilation.

For a larger hierarchy, $\kappa = 16\pi^2$, the enhancement factor is of order $T_f/T_r \sim 10^3$ and the cross-section must be $\Expect{\sigma v} \sim 10^{-23} cm^3 s^{-1}$ which can give a very good fit to PAMELA as we will show below.

Interestingly, the helicity suppression that can reduce $B_{3/2}$ (and $Y_{3/2}$) from the decay of the modulus is in general not enough to make the first factor in Eq.~(\ref{Xfrom32}) smaller than the self-annihilation abundance. For the range of numbers used in this paper, we found that the branching ratio needs to be smaller than $B_{3/2}\sim 10^{-7}$ for the first factor to be smaller while the helicity suppression of \cite{Dine} gives $10^{-4} - 10^{-5}$. This is interesting as this means that even when BBN constraints for the gravitino are evaded by having  $B_{3/2}$ small enough, one still overproduces the LSP unless the cross-section is enhanced by the factor we calculated. Of course, if the modulus mass is small and the branching ratio is suppressed because of phase space consideration, then $B_{3/2}$ can be as small as we want. In that case the non-thermal production of dark matter will be dominated by the modulus and not the gravitino. Since they would have very similar temperature of reheat, we expect about the same enhancement factor although in principle the modulus could have a very small branching ratio to dark matter (unlike the gravitino).

To summarize, the non-thermal production of dark matter from the decay of the gravitino can give rise to the correct relic density if the cross-section is larger than the canonical value by a factor of $10^3 - 10^4$. This is in the high range of what is allowed by experiments but it could be the explanation for PAMELA as we will further discuss below. If the gravitino abundance is always very small, (even smaller than what is required to satisfy BBN constraints), then we can neglect its contribution to the dark matter relic density and instead look at cosmological moduli. By tuning the mass of this modulus one can get enhancement factor between $1-10^4$. In this case one must worry that the gravitino problem is not revived in this process by ensuring $B_{3/2}$ is small enough.

\section{A Concrete Model}

In this section, we construct a successful model to implement the cosmological scenario oulined in the previous section. We study mirage mediation \cite{Choi} to a $U(1)_{B-L}$ extension of the MSSM which appears often as a typical setting from the  point of view of string phenomenology, and argue on general grounds that the dark matter is either the right handed sneutrino or the $U(1)_{B-L}$ gaugino. Further the dark matter is leptophilic since the dominant mode of annihilation is to the lightest of the new Higgs fields, and its subsequent decay mainly produces taus or muons by virtue of appropriate $B-L$ charges. The cosmological moduli and gravitino problems are addressed by the rather large cross section in such annihilation.

 In a construction of supersymmetry breaking vacua such as KKLT~\cite{Kachru:2003aw}, the volume modulus $T$ is stabilized by non-perturbative effects and then an uplifting mechanism (for example with anti-$D3$ branes) is used to obtain a supersymmetry breaking vacuum. The exponential form of the non-perturbative potential leads to a small hierarchy between the moduli mass and the SUSY breaking scale
 \beq
 m_{T} \sim m_{3/2}{\text {ln}}(M_{Pl}/m_{3/2})\; .
 \eeq
with ${\text {ln}}(M_{Pl}/m_{3/2}) \sim 4\pi^2$. If the supersymmetry breaking brane is sequestered from the visible sector, then $T$ makes $\mathcal {O}(F_T/T) \sim m_{3/2}^2/m_T $ contributions to the soft terms and one sees that the modulus contribution to soft masses is comparable to the anomaly-mediated contribution.
There is thus a natural hierarchy of sparticle, gravitino, and moduli masses given by $\mathcal {O}(4\pi^2)$. Note that in general one might expect extra moduli with masses around the SUSY breaking scale $m_\phi \sim m_{3/2}$ and while they can be used and included in the discussion we mainly discuss the minimal mirage scenario with very heavy moduli in this paper. Setting the LSP at the TeV scale, we obtain the cosmology depicted in Fig. 1.
\begin{figure}[ht]
\centering
\includegraphics[width=3.5in]{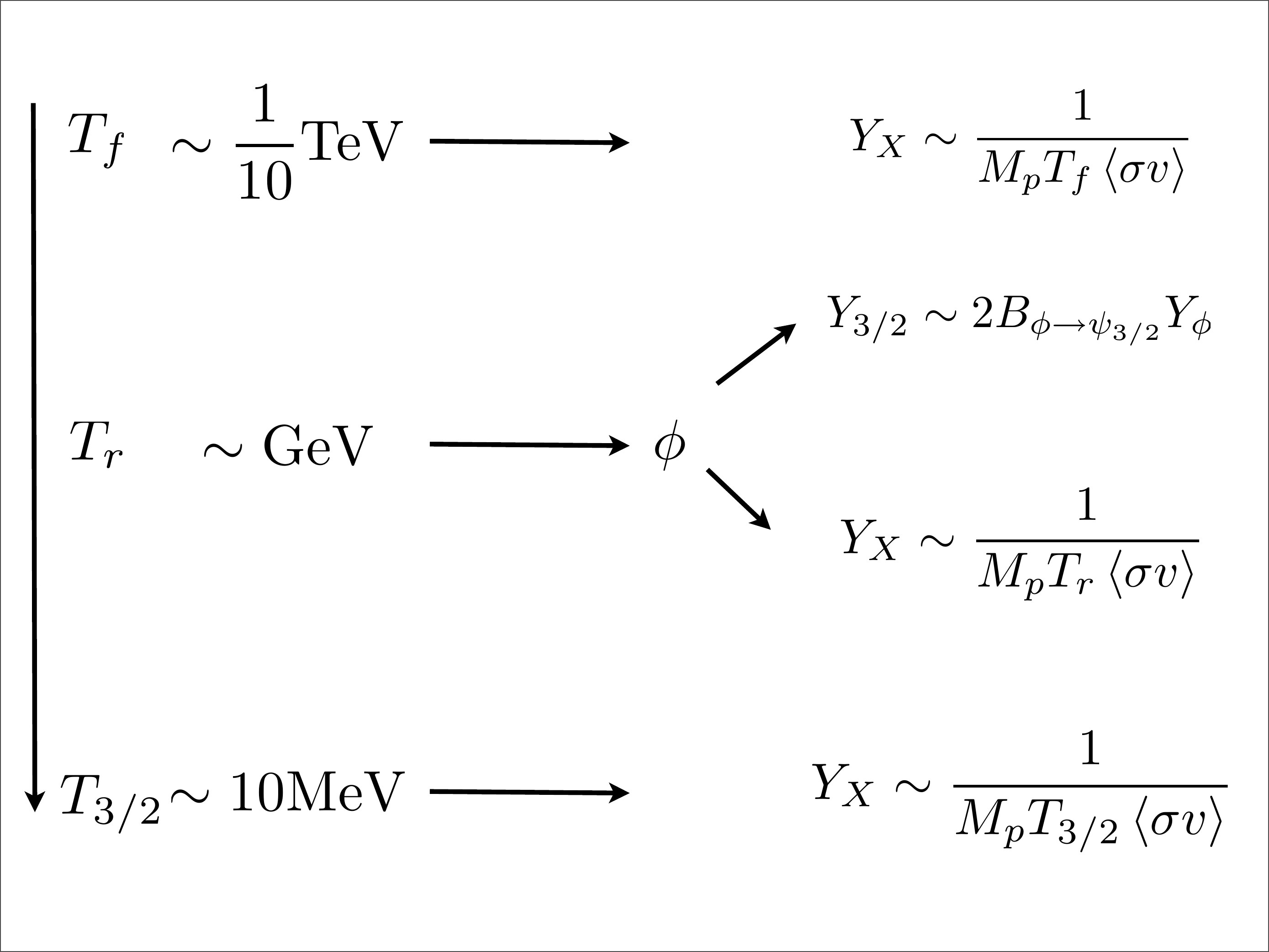}
\caption{The main thermal events in a scenario with a heavy moduli at $10^3$ TeV, a gravitino mass at $10^2$ TeV and a LSP at the TeV scale. We assume that there are no suppressed branching ratios. Enough LSP (denoted X) is produced for self-annihilation to be important and the attractor solution for the abundance is reached at each phase transition. At each reheating phase, there is entropy production and the previous abundance of dark matter is diluted (by a factor roughly of $\left(\frac{T_{new}}{T_{old}}\right)^3$ between the two phases 'new' and 'old'). The final answer in this particular set-up is to a good approximation simply given by the last decay $Y_X \sim \frac{1}{M_p T_{3/2}\Expect{\sigma v}}$.}
\end{figure}

\subsection{The MSSM with a  $B-L$ extension}

For gauge group $SU(3) \times SU(2)_L \times U(1)_Y \times U(1)_{B-L}$, $U(1)_{B-L}^3$ triangle anomaly cancellation automatically implies the existence of three right-handed (RH) neutrinos through which one can explain the neutrino masses and mixings~\cite{mohapatra}. Such extensions have been studied for a long time; this model has also been recently studied in the context of inflation~\cite{Inflation}, anomaly mediation~\cite{Kikuchi:2008gj}, dark matter~\cite{Allahverdi:2008jm,Khalil:2008ps}, and leptogenesis~\cite{Babu:2009pi}. The model contains a new gauge boson $Z^{\prime}$, two new Higgs fields $H^{\prime}_1$ and $H^{\prime}_2$, and their supersymmetric partners. The $B-L$ charge assignments are shown in Table 1. The superpotential is
\begin{equation} \label {fullsuperpotential}
W = W_{\rm MSSM} + W_N + \mu^\prime {H}^{\prime}_1 {H}^{\prime}_2 + W_{soft}
\end{equation}
where $W_N$ is the superpotential containing RH neutrinos, and $\mu^\prime$ is the new Higgs mixing parameter. Note that
\begin{equation} \label {RHsuperpotential}
W_N = (y_{D})_{ij} H_u L_i N^c_j
\end{equation}
where $y_D$ corresponds to Dirac Yukawa couplings. The new Higgs fields are neutral under MSSM charges, and do not have renormalizable couplings to quarks and leptons. The $U(1)_{B-L}$ symmetry is broken by the VEV of these new Higgs bosons, $v_1^\prime \equiv \langle H^{\prime}_1 \rangle$ and $v_2^\prime \equiv \langle H^{\prime}_2 \rangle$. The Dirac Yukawas generate small neutrino masses.

The new bosons have masses as follows. The $U(1)_{B-L}$ gauge boson $Z^{\prime}$ receives mass $m^2_{Z^{\prime}} = (27/4) g^2_{B-L} (v_1^{\prime 2} + v_2^{\prime 2})$, with $g_{B-L}$ being the $B-L$ gauge coupling. There are three physical Higgs states: the lightest $\phi$ (not to be confused with cosmological moduli which we considered in previous sections) has mass $m^2_{\phi} < m^2_{Z^{\prime}} \cos^2 2 \beta^{\prime}$, which implies $m_{\phi} \ll m_{Z^{\prime}}$ for $\tan \beta^{\prime} \equiv v_2^{\prime} / v_1^{\prime} \approx 1$. The other two Higgs states, $\Phi$ and ${\cal A}$, are heavy and have masses comparable to $m_{Z^{\prime}}$.

 If our new Higgs have charges 1 and -1, we can allow couplings: $f_{ij} H^{\prime}_1 N^c_i N^c_j$. In this scenario, the Majorana masses for RH neutrinos develop through $f_{ij}$, once $v_1^\prime \neq 0$. If we consider the case of $B-L$ breaking at the TeV scale, then, the see-saw mechanism to generate small neutrino masses  requires
 $\frac {y_D^2}{f}\frac{v_{weak}^2}{v_{B-L}} \sim 0.1$ eV, which gives $ y_D^2 \sim 10^{-10}f$. We further note that $f \sim 0.2$ (assuming $f$ to be  the largest Majorana coupling) is large enough for radiative $U(1)_{B-L}$ breaking, and small enough to guarantee that one-loop corrections to $m_{\phi}$ do not dominate over its tree-level bound. For this value of $f$, one obtains $m_{\phi} < 20$ GeV which is needed to satisfy the anti-proton data from the PAMELA experiment.

\begin{table}[tbp]
\center
\begin{tabular}{|c||c||c|c|c|c|c|c|c|c|}\hline
{\rm Fields} & $Q$ & $Q^c$ & $L$ & $L^c$ & $N$ & $N^c$ & $H^{\prime}_1$ & $H^{\prime}_2$ \\ \hline
$Q_{B-L}$ & 1/6 & -1/6 & -1/2 & 1/2 & -1/2 & 1/2 & 3/2 & -3/2 \\ \hline
\end{tabular}
\caption{The $B-L$ charges of the fields. Here $Q$, $L$ and $N$ represent quarks, leptons, and RH neutrinos respectively; while $H^{\prime}_1$ and $H^{\prime}_2$ are the two new Higgs fields. The MSSM Higgs fields have zero $B-L$ charges and are not shown in the table.}
\end{table}

We can have either $\tilde{Z^{\prime}}$ or the right-handed sneutrino $\tilde N$ as the LSP. The masses of the new Higgsinos are determined by $\mu^{\prime}$ and they are assumed to be heavy. The lightest neutralino in the $B-L$ sector lies predominantly in the  $\tilde{Z^{\prime}}$ direction. 

The beta function coefficients for the $B-L$ model are given by
\begin{equation}  \label{betafunction}
(b_{B-L}, \, b_1, \, b_2, \, b_3) = (51/4, \, 33/5, \, 1, \, -3).
\end{equation}
At the electroweak scale $g_{B-L} \sim 0.4$, in order to achieve grand unification. Note that a normalization factor of $\sqrt{3/2}$ has been used for the $B-L$ charges. The anomalous dimensions for the matter superfields (taking $y_D$ much smaller than the MSSM Yukawas) are given by
\begin{eqnarray}  \label{anomalousdimensions}
\gamma_{Q_a} &=& \frac{8}{3} g_3^2 + \frac{3}{2} g_2^2
                +\frac{1}{30} g_1^2 + \frac{1}{12} g_{B-L}^2 - (y_t^2 + y_b^2) \delta_{3a}, \nonumber \\
\gamma_{U_a} &=& \frac{8}{3} g_3^2  + \frac{8}{15} g_1^2 + \frac{1}{12} g_{B-L}^2
                - 2 y_t^2 \delta_{3a}, \nonumber \\
\gamma_{D_a} &=& \frac{8}{3} g_3^2  + \frac{8}{15} g_1^2 + \frac{1}{12} g_{B-L}^2
                - 2 y_b^2 \delta_{3a}, \nonumber \\
\gamma_{L_a} &=& \frac{3}{2} g_2^2 + \frac{3}{10} g_1^2 + \frac{3}{4} g_{B-L}^2
                - y_\tau^2 \delta_{3a}, \nonumber \\
\gamma_{E_a} &=& \frac{6}{5} g_1^2 + \frac{3}{4} g_{B-L}^2 - 2 y_\tau^2 \delta_{3a}, \nonumber \\
\gamma_{N^c_a} &=& \frac{3}{4} g_{B-L}^2  - 2 y_D^2 \delta_{3a}, \nonumber \\
\gamma_{H_u} &=& \frac{3}{2}g_2^2+\frac{3}{10}g_1^2 -3y_t^2, \nonumber \\
\gamma_{H_d} &=& \frac{3}{2}g_2^2+\frac{3}{10}g_1^2 - 3 y_b^2 -  y_\tau^2 \nonumber \\
\gamma_{H^{\prime}_1} &=& \frac{27}{4} g_{B-L}^2 , \nonumber \\
\gamma_{H^{\prime}_1} &=& \frac{27}{4} g_{B-L}^2
\end{eqnarray}
If we introduce the Majorana couplings $f$ in the model then $\gamma_{N^c_a}$ and $\gamma_{H^{\prime}}$ will get  $-f^2$ contributions in the above equation. We are not including $f$ in our analysis just for simplification and even if we include these couplings, the overall conclusion remains unchanged.
\subsection{Mirage Mediation}

Mirage mediation is a mixture of modulus and anomaly mediation. In this scheme of mediation, the gaugino and scalar masses unify at an intermediate scale (called mirage scale) below the GUT scale. This scheme occurs quite naturally in warped compactification of string theory such as in \cite{Kachru:2003aw} but we will not rely on any specific string theory construction.  The soft parameters at the GUT scale are given by
\begin{eqnarray} \label{softGUT}
M_a&=& M_0 +\frac{m_{3/2}}{16\pi^2}\,b_ag_a^2,
\nonumber \\
A_{ijk}&=&\tilde{A}_{ijk}-
\frac{m_{3/2}}{16\pi^2}\,(\gamma_i+\gamma_j+\gamma_k),
\nonumber \\
m_i^2&=& \tilde{m}_i^2-\frac{m_{3/2}}{16\pi^2}M_0\,\theta_i
-\left(\frac{m_{3/2}}{16\pi^2}\right)^2\dot{\gamma}_i
\end{eqnarray}
where $M_0, \tilde{A}_{ijk},$ and $\tilde{m}_i$ are pure modulus contributions, given as functions of the modulus $T$. Our conventions are
\begin{eqnarray}
b_a&=&-3{\rm tr}\left(T_a^2({\rm Adj})\right)
        +\sum_i {\rm tr}\left(T^2_a(\phi_i)\right),
\nonumber \\
\gamma_i&=&2\sum_a g^2_a C^a_2(\phi_i)-\frac{1}{2}\sum_{jk}|y_{ijk}|^2,
\nonumber \\
\dot{\gamma}_i&=&8\pi^2\frac{d\gamma_i}{d\ln\mu},\nonumber \\
\theta_i&=& 4\sum_a g^2_a C^a_2(\phi_i)-\sum_{jk}|y_{ijk}|^2
\frac{\tilde{A}_{ijk}}{M_0},
 \end{eqnarray}
where the quadratic Casimir $C^a_2(\phi_i)=(N^2-1)/2N$ for a fundamental representation $\phi_i$ of the gauge group $SU(N)$, $C_2^a(\phi_i)=q_i^2$ for the $U(1)$ charge $q_i$ of $\phi_i$, and $\sum_{kl}y_{ikl}y^*_{jkl}$ is assumed to be diagonal. To set input parameters, we define the ratios
\begin{equation}
\alpha\,\equiv\,\frac{m_{3/2}}{M_0\ln(M_{Pl}/m_{3/2})},\quad
 a_i\,\equiv\,\frac{\tilde{A}_{i}}{M_0}, \quad
 c_i\,\equiv\, \frac{\tilde{m}_i^2}{M_0^2},
 \end{equation}
where $\alpha$ represents the anomaly to modulus mediation ratio, while $a_{i}$ and $c_i$ parameterize the pattern of the pure modulus mediated soft masses.

The input parameters in RG running are
\begin{equation}
M_0, \, \alpha, \, a_i, \, c_i, \, {\text {tan}}\beta,
\end{equation}
where one could also choose $m_{3/2}$ as an input in place of $\alpha$.

In terms of brane constructions in type IIB, if the matter fields live on the entire worldvolume of the $D7$ from which visible sector gauge fields originate, then $a_i = c_i = 1$ while if the matter fields live on intersections of $D7$s, then $a_i = c_i = 1/2,0$~\cite{Choi}. Compactifications with dilaton-modulus mixing, realized, for example, in type IIB by the presence of gauge flux on the $D7$, can easily lead to other positive values of $\alpha, c,$ and $a$. In typical compactifications, $m_{3/2}$ is set by appropriate choice of flux contributions to the superpotential
\begin{eqnarray}
W & = & W_{\rm flux} \, + A e^{aT}
\end{eqnarray}
with ${\rm ln}(A/W_{\rm flux}) \sim 4\pi^2$ and $m_{3/2} \sim W_{\rm flux}$.

For typical values of the model parameters, the gaugino mass runnings are shown in Figure 2.

\begin{figure}[ht]
\centering
\includegraphics[width=3.5in]{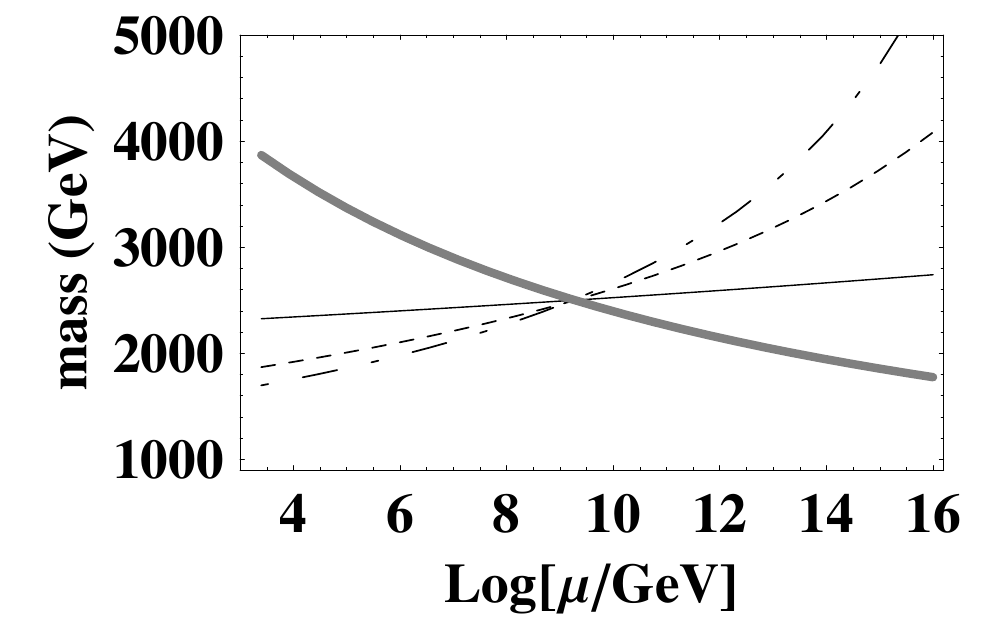}
\caption{Running of gaugino masses in the $B-L$ model. The thick gray, solid, dotted, and dashed-dotted lines are the gluino, Wino, Bino, and $\tilde{Z^{\prime}}$ respectively. We have used $a_i$=$c_i$=1, $\tan\beta=10$, $m_{3/2} = 77\,$TeV, $M_0 = 2.5\,$TeV ($\alpha = 1$)}
\end{figure}
The ratio of the gaugino masses $m_{\tilde{Z^{\prime}}} : m_{\rm Bino} : m_{\rm Wino} : m_{\rm gluino}$ is obtained as
\begin{equation}
(1 \, : 1.2\, : 1.8\, : 3.6).
\end{equation}
Evidently, the Bino is heavier than $\tilde{Z^{\prime}}$. In usual mirage mediation to the MSSM, the lightest neutralino is mostly Bino for $\alpha \leq 1$, with the Higgsino component increasing with increasing $\alpha$. This is true for various values of $a_i,c_i$, and for $\tan \beta = 10$~\cite{Choi}. This conclusion changes  in the $B-L$ extension, since the beta functions of the MSSM Yukawas get negative contributions from $g_{B-L}$, slightly raising their low energy values. This difference feeds positively into the beta function of the MSSM $\mu$ parameter, thus lowering its low energy value compared to pure MSSM. In principle, this would mean that the Higgsino component of the lightest neutralino in the MSSM sector would begin to dominate for slightly lower values of $\alpha$.

In Figure 3, we plot the RG evolution of the sfermions. For $m_{3/2} = 77\,$TeV, $M_0 = 2.5\,$TeV, and $c_i = a_i = 1$ ($\alpha =1$) one sees that the scalars are heavier than the $B-L$ gaugino. The RH sneutrino is lighter than the MSSM sfermions, due to the fact that in the case of sleptons we have contributions from MSSM gauge couplings in addition to the $U(1)_{B-L}$ gauge couplings.

We can make the right handed sneutrino even lighter by choosing $c_\nu$ appropriately. We show one such example in Figure 3 (the solid line at the bottom) and in fact, the right handed sneutrino can be the LSP of this model. For this case, we take,  $m_{3/2} \sim 200\,$TeV, $M_0 = 5\,$TeV, $c_{\nu} = a_{\nu} = 0.3$, $c_i = a_i = 1$ for all other particles ($\alpha = 1.3$), one obtains the right-handed sneutrino as the LSP with mass around $1.5\,$TeV while the gauginos and other scalars are around $\sim 3\,$TeV or so. The cross section in this case is around $\sim 10^{-23} cm^3 s^{-1}$.

So we conclude that either the $B-L$ gaugino or the right-handed sneutrino can be the LSP in this model.

\subsection{Explanation of the observed anomalies in cosmic rays}

In order to explain the recent  cosmic ray data,  we need  electron-positrons in the final states of LSP annihilation. In this model, the LSP ($\tilde{Z'}$ or the sneutrino) annihilates into light Higgs bosons (from the $B-L$ sector) which then decays into a pair of taus predominantly ~\cite{Allahverdi:2008jm}.

The taus then decay into electron-positron pair. A recent analysis of the data showed that~\cite{Allahverdi:2008jm} in order to explain the excess by using $\tau$s, we need an enhancement factor of $10^3$ for the annihilation cross-section. The annihilation cross-section does not have any p-wave suppression. The typical value of the LSP mass that fits the data in this model with this enhancement factor is about 1-2 TeV.  The LSPs annihilate to lightest Higgs  ($\phi$) of the B-L sector, whose mass is controlled by the VEVs of the new Higgs fields. For comparable VEVs, i.e. for $\tan \beta^{\prime} \approx 1$, it can be very small without any tuning of the soft masses in the Higgs sector. We can choose this mass to be between $O(1)$~GeV and 20~GeV in order to be in complete agreement with the anti-proton data. For $2 m_\tau < m_{\phi}$ the dominant decay mode is to $\tau^{-} \tau^{+}$. If we assume the $\phi$ mass to be $>$20 GeV, then the Br of $\phi\rightarrow b\bar
 b$ is about 1/7 of $\phi\rightarrow\tau\tau$ due to the B-L charges. In this case the anti-proton data is still satisfied up to a factor of 2. However, the computation of anti-proton flux involves a large theoretical uncertainty~\cite{Bergstrom:2006tk}. In our model both small and large values of $\phi$ are allowed, and as we have already discussed, a small Higgs mass requires smaller values of $\tan\beta'$.  If $m_\phi$ is slightly less than $2 m_\tau$, $\phi$ can decay either to $c {\bar c}$ or $\mu^{-} \mu^{+}$ with comparable branching ratios. It is possible to reduce the $\phi$ mass further to be below $2 m_c$, and make $\mu^{-} \mu^{+}$ final state the dominant decay mode.

We now discuss the two options for the LSP in this model.
\begin{enumerate}
\item If $\tilde{Z'}$ gaugino is the LSP, then from the cosmology analysis we find a typical enhancement $\sim 10^4$ gives the correct relic density. This value of the enhancement  follows from Eq.~(\ref{enhancement}) and the fact that the gravitino  mass and the LSP  mass are typically correlated by a hierarchy $\kappa \sim {\text {ln}}(M_{Pl}/m_{3/2}) \sim 4 \pi^2$. One can lower the enhancement by raising $M_0$, but in our case we cannot raise $M_0$ too much since we want the gaugino mass to be $1.5-2$ TeV.

An enhancement factor of $10^{4}$ can have problem with BBN~\cite{Hisano:2009rc}\footnote{The most stringent bounds coming from $^{3}$He/D give the upper bound on the cross section for non-thermal dark matter annihilating to tau to be $\sim 3 \times 10^{-23} cm^3 s^{-1}$.}, where it is claimed that the enhancement factor should be less than $10^{2-3}$. If this enhancement factor is somehow accommodated by the BBN data, then  it is possible to generate such a large cross-section by having an annihilation funnel of $B-L$ gauginos into a pair of the $\phi$, the lightest boson in the $B-L$ Higgs sector via the $s$-channel exchange of the $\phi,~\Phi$ (heavy Higgs). The S-channel resonance of this process enhances the cross-section to the required value. While a $1.5$ TeV LSP has been fitted with PAMELA data for an enhancement $10^3$, we note that astrophysical uncertainties (for example, a choice of isothermal DM density profile instead of NFW in the halo function can easily give rise to a factor 2-5 uncertainty, and a factor of $2$ uncertainty in the energy loss coefficient of positrons \cite{Cirelli:2009uv}) could provide a fit with enhancement $10^4$.

\item The best option for the LSP is the new sneutrino ($\tilde N$). In this case, the hierarchy between the gauginos and the gravitino remains $\sim 4\pi^2$, but for small values of $c_\nu$ in the mirage mediation input parameters, the sneutrino can be made much lighter than the gauginos. Thus, the hierarchy between the sneutrino LSP and the gravitino becomes $\kappa \sim 16\pi^2$ and a lower cosmological enhancement of $10^3$ (hence annihilation cross-section around $10^{-23}cm^3 s^{-1}$) is obtained from Eq.~(\ref{enhancement}). An enhancement factor of $10^3$ allows us to fit the  PAMELA data. The BBN bound of \cite{Hisano:2009rc} is also satisfied without any difficulty. On the particle physics side,  this cross-section can be obtained with and without the heavy Higgs annihilation funnel. The annihilation amplitude is proportional to the gauge boson mass which appears in the ${\rm{\tilde N}}^* {\rm{\tilde N}}\phi$ vertex \footnote{This vertex arises from
$
V_D \supset \frac{1}{2} D^2_{B-L}$,
where
$
D_{B-L} = \frac{1}{2} g_{B-L} \left[Q_{1} ({\vert H^{\prime}_1 \vert}^2 - {\vert H^{\prime}_2 \vert}^2) + \frac{1}{2} {\vert \tilde N \vert}^2 + ... \right] .
$. $H_i^{\prime}$ are  new Higgs~\cite{Allahverdi:2008jm}}.
The dominant channel is ${\rm {\tilde N}}^* {\rm{\tilde N}} \rightarrow \phi \phi$ via the $s$-channel exchange of the $\phi,~\Phi$, the $t$,~$u$-channel exchange $\tilde N$, and the contact term $|\tilde N|^2\phi^2$. The $s$-channel $Z^\prime$ exchange is subdominant because of the large $Z^\prime$ mass (as required by the experimental bound on $m_{Z^\prime}$).
The sneutrino annihilation into $\nu \bar{\nu}$ final states is at least an order of magnitude below the $\phi \phi$ final states. Other fermion final states, through $s$-channel $Z^\prime$ exchange, have even smaller branching ratios (these fermion-anti-fermion final states are p-wave suppressed).

Since the cosmological enhancement is sufficient to explain the PAMELA data, we do not need any enhancement due to Sommerfeld effect. Sommerfeld enhancement requires~\cite{Allahverdi:2008jm} the model parameters to be tuned very accurately, and this can be easily prevented.
\end{enumerate}
\begin{figure}[ht]
\centering
\includegraphics[width=3.5in]{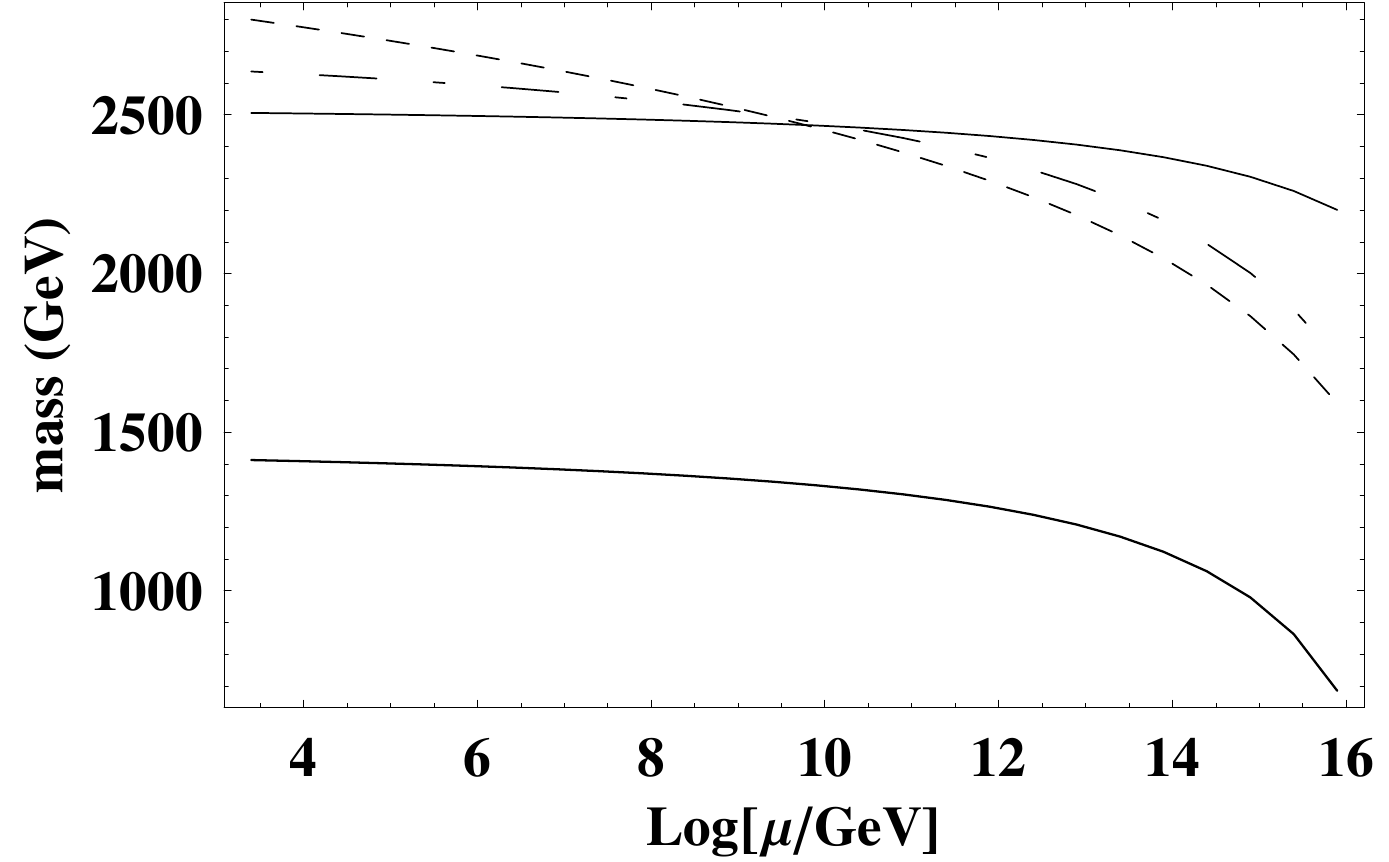}
\caption{Running of right sneutrino ($\tilde N$), left selectron and right selectron masses in the $B-L$ model. From top, the  solid, dotted, and dashed-dotted lines are the sneutrino, left selectron and right selectron masses. respectively. We have used $a_i$=$c_i$=1, $\tan\beta=10$, $m_{3/2} = 77\,$TeV, $M_0 = 2.5\,$TeV ($\alpha = 1$). The solid line which does not go through the mirage point corresponds to $\tilde N$ for $c_\nu$=0.3. }
\end{figure}

One interesting aspect of sneutrino LSP is that they can be probed in direct detection experiments~\cite{Allahverdi:2008jm}. The direct detection cross-section for sneutrino-nucleon scattering is mediated by $Z'$ exchange and the cross-section can be quite large. To explain the PAMELA data in our model we need to use large  $Z'$ gauge boson mass (if we do not use the heavy Higgs annihilation funnel) and consequently the direct detection cross-section is reduced. It is interesting to note that if instead we use thermal dark matter with Sommerfeld enhancement in our model, a smaller $Z'$ mass is needed to explain the dark matter content (again, if we do not use the Higgs annihilation funnel). Thus the direct detection cross-section in the case of thermal dark matter is much larger. We show this feature in Figure 4.

Therefore, combining the direct detection result with PAMELA results it is possible to distinguish the cosmological enhancement from the Sommerfeld enhancement. If we choose the annihilation funnel to satisfy the dark matter content, we can allow smaller values of $Z^\prime$ and the direct detection cross-section can be larger.
\begin{figure}[ht]
\centering
\includegraphics[width=3.5in]{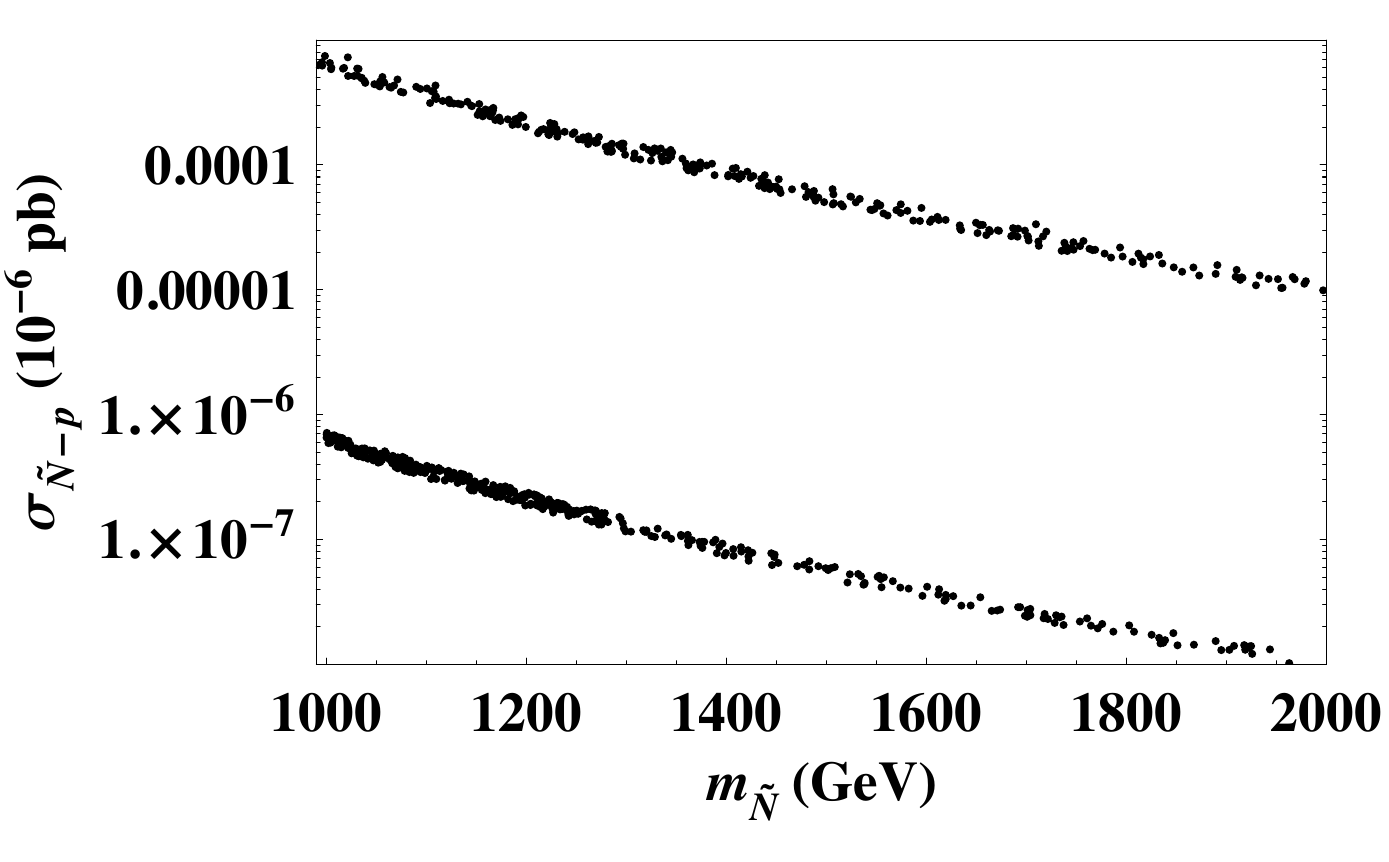}
\caption{Sneutrino-nucleon scattering cross-section as a function of sneutrino mass. The top line corresponds to the thermal case where we need Sommerfeld enhancement to explain the PAMELA data. The bottom line corresponds to the non-thermal case as discussed in this work. The correct relic density is satisfied for both cases.}
\end{figure}

\section{Discussion}

In this paper, we have studied non-thermal dark matter, the moduli and gravitino problem in the light of the PAMELA data. As a phenomenological model, we have considered a $U(1)_{B-L}$ extension of the MSSM where supersymmetry breaking is mediated by mirage mediation.

We have found that the final decaying particle that non-thermally produces LSP may be either a cosmological modulus or the gravitino. Cosmological moduli typically produce gravitino, and the decay of either at a temperature above BBN tends to overproduce dark matter. A larger annihilation cross section for dark matter can naturally ease this overproduction problem.

We have shown that it is possible to solve the moduli/gravitino problem in the $B-L$ model with mirage mediation. The natural hierarchy between LSP and gravitino/moduli in mirage mediation allows the gravitino to decay above BBN, while maintaining an LSP in the $1-2$ TeV range. Moreover, a large enhancement of the  annihilation cross section of $10^{3-4}$ is readily obtained in this model, which solves the overproduction problem and can fit the PAMELA data.

The LSP can be either the new $B-L$ gaugino or the right-handed sneutrino. Both of these annihilate to the light Higgs of the new sector. This Higgs primarily decay into tau for $m_\phi>m(2\tau)$ due to the B-L charges. For the $B-L$ gaugino LSP $\sim 1.5$ TeV, an enhancement $\sim 10^4$ is calculated cosmologically from the decay of the gravitino. This enhancement does not do well with BBN constraints.
For the sneutrino LSP $\sim 1.5$ TeV, a cosmological enhancement of $10^3$ is calculated from a larger hierarchy between LSP and gravitino. This can be obtained by an appropriate choice of mirage mediation parameters. This enhancement explains the recently observed anomalies in cosmic rays and is allowed by the BBN constraint.
The sneutrino LSP has interesting consequences for direct detection experiments. In fact, in this case it is possible to distinguish between models of non-thermal dark matter and models with thermal dark matter that utilize Sommerfeld enhancement in certain regions of parameter space.

\section{Acknowledgement}
The authors would like to thank P. Kumar, Y. Santoso, D. Shih, and S. Watson for useful discussions.
The works of BD, L.L and K.S. are supported in parts by DOE grant DE-FG02-95ER40917 and NSF grant PHY-0505757,  PHY05-51164.

\end{document}